\documentclass[reprint,superscriptaddress,nofootinbib]{revtex4-2}

\usepackage{hyperref, multirow, color, bbold, textcomp}
\usepackage[dvipsnames]{xcolor}
\usepackage{orcidlink}
\usepackage[ruled]{algorithm2e}
\usepackage{graphicx}
\usepackage[caption=false]{subfig} %

\usepackage{tikz, varwidth, changepage}
\usetikzlibrary{calc,positioning,shapes,shapes.geometric,shapes.misc,
  shadows,arrows,arrows.meta,math,decorations.pathreplacing}

\tikzstyle{startstop} = [rectangle, rounded corners, minimum width=3cm,
minimum height=1cm,text centered, draw=black, fill=red!30]
\tikzstyle{io} = [trapezium, trapezium left angle=70, trapezium right angle=110,
minimum width=3cm, minimum height=1cm, text centered, draw=black, fill=blue!30] 
\tikzstyle{decision} = [diamond, minimum width=3cm, minimum height=1cm, text
centered, draw=black, fill=green!30]
\tikzstyle{arrow} = [thick,->,>=stealth]

\newcommand{\spectre}{\texttt{SpECTRE}}
\newcommand{\spec}{\texttt{SpEC}}

\usepackage{amsthm}
\newtheorem{statement}{Statement}

\usepackage{amsmath}

\usepackage[normalem]{ulem} %

\newcommand{\CornellPhysics}{Department of Physics, Cornell University, Ithaca,
  New York 14853, USA} 

\newcommand{\CornellLepp}{Laboratory for elementary Particle Physics, Cornell
  University, Ithaca, New York 14853, USA} 

\newcommand{\CornellCcaps}{Cornell Center for Astrophysics and Planetary
  Science, Cornell University, Ithaca, New York 14853, USA}

\newcommand{\Caltech}{Theoretical Astrophysics 350-17, California Institute of
  Technology, Pasadena, CA 91125, USA}

\begin{document}

\title{Horizon tracking for asynchronous parallel black hole simulations}

\author{Kyle C.~Nelli\orcidlink{0000-0003-2426-8768}}
 \email{knelli@caltech.edu}
 \affiliation{\Caltech}
\author{William Throwe\orcidlink{0000-0001-5059-4378}}
 \affiliation{\CornellCcaps}
\author{Nils Deppe\orcidlink{0000-0003-4557-4115}}
 \affiliation{\CornellPhysics}
 \affiliation{\CornellLepp}
 \affiliation{\CornellCcaps}
\author{Mark A.~Scheel\orcidlink{0000-0001-6656-9134}}
 \affiliation{\Caltech} 
\author{Lawrence E.~Kidder\orcidlink{0000-0001-5392-7342}}
 \affiliation{\CornellCcaps} 
\author{Nils L.~Vu\orcidlink{0000-0002-5767-3949}}
 \affiliation{\Caltech}
\author{Saul A.~Teukolsky\orcidlink{0000-0001-9765-4526}}
 \affiliation{\CornellCcaps} 
 \affiliation{\Caltech}

\begin{abstract}

  In the field of gravitational wave science, next-generation detectors will be
  substantially more accurate than the current suite of detectors. Numerical
  relativity simulations of binary black hole (BBH) gravitational waveforms must
  become faster, more efficient, and more accurate to be used in analyses of
  these next-generation detections. One approach, which the \spectre~code
  employs, is using spectral methods for accuracy along with asynchronous
  task-based parallelism to avoid idle time in simulations and make the most
  efficient use of computational resources. When writing an asynchronous
  application, algorithms must be redesigned compared to their synchronous
  counterparts. To illustrate this process, we present novel methods for
  dynamically tracking the apparent horizons in evolutions of BBH mergers using
  a feedback control system, all in the context of asynchronous parallelism. We
  also briefly detail how these methods can be applied to binary neutron star
  simulations performed with asynchronous parallelism.

\end{abstract}

\maketitle

\section{Introduction\label{sec:dgscl introduction}}

Gravitational waves provide a new window into our universe. In order to observe
these gravitational waves, detectors such as LIGO~\cite{LIGOScientific:2014pky},
Virgo~\cite{VIRGO:2014yos}, KAGRA~\cite{2021PTEP.2021eA101A}, and
GEO600~\cite{Dooley:2015fpa} have been built and in 2015, the first
gravitational waves from the merger of a binary black hole (BBH) system were
observed~\cite{PhysRevLett.116.061102}. Since the first detection, there have
been several hundred detections of compact binary systems, including BBHs,
binary neutron stars, and binary black-hole neutron-star
systems~\cite{Capote:2024rmo}. To compare these observations to theory, one must
solve Einstein's equations of general relativity using numerical relativity
(NR), which is needed because of the lack of a closed-form solution to the
two-body problem in general relativity. A number of different NR codes have been
developed to simulate compact object mergers. Furthermore, several publicly
available catalogs of gravitational waveforms have been released for use in
comparing to observations~\cite{Mroue:2013xna, Jani:2016wkt, Healy:2017psd,
Healy:2019jyf, Boyle:2019kee, 2025sxscatalog, Healy:2020vre, Healy:2022wdn,
Ferguson:2023vta}.

In order to expand the number of detections and to perform more accurate
observations, there are plans to build ``next-generation'' gravitational
wave detectors such as the Laser Interferometer Space Antenna
(LISA)~\cite{2017arXiv170200786A}, the Einstein
Telescope~\cite{2010CQGra..27s4002P}, and Cosmic Explorer~\cite{Evans:2021gyd}.
While the current NR catalogs of gravitational waveforms are sufficient for
understanding observations from the current set of detectors, they are
inadequate for analyzing observations from these next-generation detectors.
The gravitational waveforms that are needed must be at least $10\times$ more
accurate than the current catalogs~\cite{Purrer:2019jcp, Ferguson:2020xnm,
Jan:2023raq}.

To solve this problem, the NR community has been developing new or
improved codes to meet
the accuracy requirements for next-generation detectors. Some of these codes are
\spec~\cite{SpECwebsite}, \spectre~\cite{spectrecode_2025},
AthenaK~\cite{Zhu:2024utz}, and codes using the Einstein Toolkit with the
CarpetX driver such as GRaM-X~\cite{Shankar:2022ful} and
AsterX~\cite{Kalinani:2024rbk}. The approaches of these new codes can be split
into two broad categories: 1) Finite-difference methods run on GPUs and 2)
Spectral methods run on CPUs. Most codes fall into the first category, taking
advantage of the power of GPUs to perform a large number of computations very
quickly using large chunks of the computational domain. Finite difference
methods are very robust, making them appealing for the long simulations needed
in numerical relativity. Finite difference methods typically use a
cartesian grid, which lends itself very well to GPUs. The new
CarpetX driver implemented in the Einstein Toolkit is built on top of the
AMReX~\cite{amrex} adaptive mesh refinement infrastructure. AthenaK makes use of
the Kokkos framework~\cite{kokkos} for performance-portable simulations so the
user can focus on the equations rather than the parallel infrastructure.

A few codes, such as \spectre, \spec, and \texttt{NMesh}~\cite{Tichy:2022hpa},
fall into the second category and take advantage of the computational efficiency
of spectral methods over finite-difference methods. \spectre~is distinct from
\spec~and \texttt{NMesh} because it uses asynchronous parallelism, while the
others use some form of synchronous parallelism. The goal of asynchronous
parallelism is to have as little
idle time as possible, no matter the number of resources you run on.

In this work we detail the parallel algorithms implemented in \spectre~that
efficiently make use of computational resources on CPUs. The paper is structured
as follows: In section \ref{sec:async} we introduce the concept of asynchronous
parallelism and how it differs from the synchronous case. We illustrate with a
straightforward example the difference between the two using input/output (I/O).
In section \ref{sec:horizon-finder}, we detail how to implement a horizon finder
in the context of asynchronous parallelism. In section
\ref{sec:sync-control-system}, we introduce the synchronous aspects of the
feedback control system in \spectre\ along with some necessary context about the
domain and time-dependent coordinate mappings we use. Then in section
\ref{sec:async-control-system} we detail changes necessary to implement the
feedback control system with asynchronous parallelism, along with a number of
subtleties that arise. In section \ref{sec:bns} we briefly explain how the
feedback control system can be applied to binary neutron star simulations done
with asynchronous parallelism in \spectre. And finally, we summarize and
conclude our findings  in section \ref{sec:conclusion}.

\section{Asynchronous parallelism}
In this section, we first introduce the concept of asynchronous
parallelism and how it differs from a synchronous application. It is important
to make the differences explicit as they will shape how we think about and
structure our asynchronous algorithm.

\subsection{Difference from synchronous parallelism}\label{sec:async}

Synchronous parallel applications are typically implemented using the
widely-used Message Passing Interface (MPI), where the problem is divided as
evenly as possible among MPI ranks. Typically, there is one rank per
core, with each rank running the same application in parallel. For all
but the most computationally-trivial problems, the ranks are not completely
independent. Instead, ranks must receive messages from each other. This is
accomplished using the various \texttt{MPI\_Send/MPI\_Receive} calls to send and
receive messages between the MPI ranks at specific pre-determined stages during
the computation. Also, each rank typically pauses at each of these
message-exchange stages, continuing execution only after all the ranks have
exchanged messages.

Another method for implementing synchronous parallel applications is to instead
have only one MPI rank per node. The rest of the cores on the node are then used
for OpenMP acceleration in \texttt{for} loops. While this does differ from the
typical MPI application because not every core runs the same application,
there's still one core on a node which runs the same application and is strongly
synchronized between other ranks. Therefore, this method is still synchronous,
but may have better performance.

Asynchronous parallelism (also called task-based parallelism or asynchronous
many-task parallelism) on the other hand, is a type of parallelism that makes
use of small ``tasks'' that are launched on any available resource from a master
process. A task can be thought of as a function that runs on some resource, can
edit data that is stored in memory, and usually doesn't return a value. Several
third-party libraries such as Uintah~\cite{10.5555/822085.823309},
Chapel~\cite{10.5555/2753024.2753030},
Charm++~\cite{laxmikant_kale_2020_3972617}, Kokkos~\cite{kokkos},
Legion~\cite{7529975}, and
PaRSEC~\cite{chevalier2019protocolasynchronousreliablesecure} implement
task-based parallelism in slightly different ways. Often asynchronous
parallelism is implemented utilizing MPI with threading and asynchronous
send/receive calls. Therefore, the ``driver process'' is typically an MPI rank
and the ``available resources'' are the threads managed by that MPI rank. In our
application \spectre~we choose to use Charm++, but we are not locked into that
choice. The methods presented here will be applicable to any asynchronous
runtime system.

One of the critical differences between asynchronous and synchronous
applications is the guarantee of message ordering. According to the MPI
standard~\cite{mpi50}, most synchronous messages are \textit{non-overtaking},
meaning that if a single sender sends two messages A and B in order to the same
destination, they will be received in the same order (A then B). Non-overtaking
messages allow for deterministic message ordering in an application and a
guarantee of synchronicity. In asynchronous runtime systems, however, this
relatively simple requirement is no longer guaranteed. This fact is fundamental
enough to the following algorithms to give it a formal statement:

\begin{statement}
  \label{statement:a_before_b}
  If a single sender sends successive messages A then B to the same receiver,
  there is \textbf{no} guarantee which message will arrive first at the
  receiver.
\end{statement}

The consequences of statement \ref{statement:a_before_b} are subtle, yet
cascading. Applications that were originally written to be synchronous must
typically be completely rewritten to make them suitable for an asynchronous
algorithm. Communication, I/O, and global synchronizations must all be
rethought.

Lastly, we define some terminology that was used in this section and will be
reused throughout this paper. As stated before, a \textit{task} can be thought
of as a function that typically doesn't return a value. This task will typically
be run on a \textit{thread}, a process of the application, and that thread will
be assigned to a \textit{core} of a CPU. For this reason, we typically use
thread and core interchangeably even though they are technically different. We
then have a \textit{node} which is a collection of cores. Finally, we'll define
the concept of an \textit{algorithm} to be a collection of tasks and associated
data stored in memory that these tasks will have access to. An example of an
algorithm is described in the next section.

\subsection{I/O Example\label{sec:io}}

To give a straightforward example of an algorithm that needs to be rethought
when using task-based parallelism, let's consider I/O from our computation.
Imagine that we have data distributed across our computational resources that
depends on simulation time. We would like to write the data to a file on disk at
each simulation time. (For simplicity, in the remainder of this paper we will
use ``time'' to refer to simulation time. When we need to refer to wallclock
time, we will do so explicitly.) Suppose that one MPI rank is dedicated to
writing the data to disk. In a synchronous implementation (we'll use \spec~as an
example), each rank would compute the necessary data to write to disk at a given
time and wait for all other ranks to finish at that same time. Then, all ranks
send their data synchronously to the dedicated writing rank, the data is
combined, and written out to disk. Once writing is complete, each rank can
continue the application and advance to the next time, combine the new data, and
write it to disk. The drawback to this method is the scenario when a rank
finishes its work quickly and could potentially move along to the next time and
send data for I/O at this new time, but can't because it must wait to
participate in the first time's reduction.

Now we consider I/O in an asynchronous implementation with the same setup,
replacing ranks with tasks and the writing rank with a passive thread that will
run tasks which actually write the data to disk. All tasks start computing their
data that will be written to disk, and some finish more quickly than others.
Because this is an asynchronous implementation, a task that finishes quickly
doesn't need to wait for other tasks to finish. It sends its own portion of the
data and launches a task on the dedicated writing thread and is now free to
continue doing work. This freedom to keep doing additional work without
synchronizing with any other tasks can improve efficiency, but must be handled
carefully because race conditions could arise.

Consider a case where tasks run on a single node and some tasks finish faster
than others. A direct example of Statement \ref{statement:a_before_b} would be
that a task A is able to do so much work that it sends two separate messages to
the dedicated writing thread for different times before task B even sends its first
message. This possibility implies that an I/O algorithm must not only be able to
accept data from tasks at a single time in any order, but should also be able to
accept data at any time because it may arrive ``out of order.'' This will add
code complexity to an I/O algorithm and can (and will) inform how data is stored
in memory and on disk to accommodate receiving portions of the data out of
temporal order. We don't detail a specific implementation on how to store this
data because it is application dependent, but we do suggest that the solution
involves 1) knowing how many tasks will be sending data to the writing thread
so the I/O algorithm knows when it has received all data and 2) storing the data
in a map-like data structure sorted by time.

\subsection{Complications with I/O example\label{sec:io-complications}}

It is worth noting the potential issues that can arise if an asynchronous
algorithm is implemented incorrectly, as they can cause issues that one may not
expect. The I/O example is a passive asynchronous algorithm, meaning that there
is no feedback from the writing thread onto the rest of the application, so we
don't have to worry about deadlocks (c.f. section~\ref{sec:deadlock}) here.
However, an issue can arise if the number of tasks contributing data to I/O at
each time is smaller than the number of messages the writing thread expects to
receive. In this scenario, data will continue to be sent to the writing thread,
but it will never be written to disk and purged from memory, causing a ``memory
leak''. ``Memory leak'' is in quotes because it is not the typical memory leak
found in computer science where a pointer's memory is not freed correctly, but
is rather an algorithmic bug that causes data to be stored for too long. This
can eventually result in an out-of-memory (OOM) error.

Another possible issue can occur even if the I/O algorithm is implemented
correctly. Consider a case where some tasks are able to advance far ahead in time
and continually send data to the writing thread, while other tasks lag behind
and don't send any data (possibly their workload is much larger). A similar OOM
issue to the previous one can arise where the writing thread keeps storing more
and more data without writing to disk since it has not received data from all
tasks at a given time. However, now the OOM issue is not due to a bug, but is
due to the (possibly random) unequal workload among tasks. This must be dealt
with in some way, possibly by limiting how far ahead in time tasks can proceed,
or having the writing thread write whatever partial data it has to disk when it
reaches some threshold memory usage.

\section{Apparent Horizon finder\label{sec:horizon-finder}}

We start by detailing how a standalone apparent horizon finding algorithm can be
implemented in an asynchronous application. Specifically, we'll describe what is
implemented in \spectre. For the synchronous part of the horizon finding
algorithm, we use a fast-flow method similar to \cite{Gundlach:1997us}. This is
an iterative method that takes as input a trial surface, and at each iteration
it produces a new trial surface that is closer to the actual apparent horizon.
However, the details of the algorithm are unimportant for the discussion here as
it does not affect the parallelism. What is important is that for each iteration
of the method, evolution variables in the volume (for example the spatial metric
and extrinsic curvature) must be interpolated onto the trial surface. The volume
that is evolved is usually split into smaller partitions we'll call cells. For a
synchronous implementation, the horizon finding algorithm is relatively
straightforward:

\begin{enumerate}
\item \label{lst:horizon-finder-1} All cells finish their time step. Evolution
variables are now in a consistent state at the same time.
\item \label{lst:horizon-finder-2} Broadcast the points of the trial surface to
all cells. If a cell contains trial surface points, interpolate evolution
variables to those points.
\item Send the interpolated evolution variables on the trail surface points to
a single process.
\item Perform one iteration of the fast-flow algorithm. Get a new trial surface
(or finish if the algorithm converges and an apparent horizon is found).
\item Go to step \ref{lst:horizon-finder-2}.
\end{enumerate}

The difficulty of implementing a horizon finder with task-based parallelism
arises in steps \ref{lst:horizon-finder-1} and \ref{lst:horizon-finder-2} above.
As mentioned before in the I/O example (section \ref{sec:io}), cells are free to
evolve as far into the future as they can regardless of what other cells may be
doing (so long as a cell has sufficient data from neighboring cells). This
poses a significant problem because the iterative horizon finding algorithm
requires the evolution variables to persist at a given time long
enough for all iterations to finish and for the algorithm to converge. We could
easily solve this like we did for the synchronous case where all cells stop at a
given time until the horizon-find is finished. However, this defeats
the advantages of asynchronous parallelism. The main quandary is then how to
store the evolution variables from cells while also allowing those cells to
evolve past the time a horizon-find is supposed to happen. A secondary concern
is what to do if a cell evolves past \textit{multiple} horizon-find times.

Our solution is to have a passive thread, called the horizon interpolator, whose
entire job is to receive volume evolution variables from cells, store and sort
them by time, receive a trial surface at a given time, and determine if enough
cells have sent their evolution variables at that time to complete the
interpolation. If so, then the interpolation is done, and one iteration of the
fast-flow algorithm is performed. Once an iteration is finished and a new trial
surface is found, the horizon interpolator is once again asked to interpolate
onto the new trial surface.

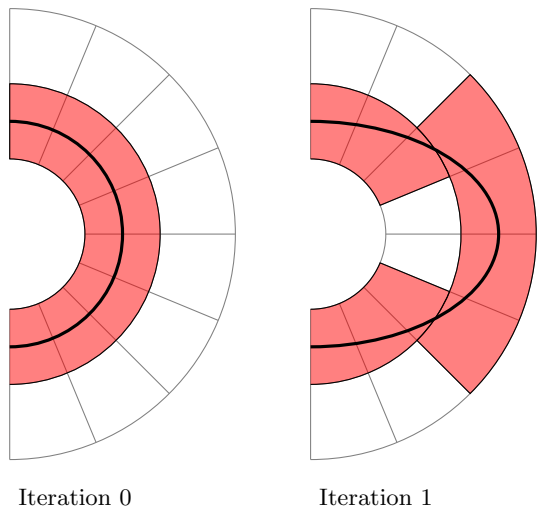
\begin{figure}
\begin{tikzpicture}
\tikzmath{
  \lc = -1.5;
  \rc = 2.5;
  \numrays = 8;
  \numradii = 3;
  \maxradius = 3;
  \firstradius = \maxradius/\numradii;
  \arc1 = 1.5*\firstradius;
  \arc2 = 2.5*\firstradius;
}
\foreach \s in {1,...,\numradii} {
  \draw[gray] (\lc,-\s*\firstradius) arc(-90:90:\s*\firstradius);
}
\foreach \ang in {0,...,\numrays} {
  \draw[gray] (\lc,0)+(90 - \ang*180/\numrays:\firstradius) -- ++(90 - \ang*180/\numrays:\maxradius);
}
\draw[fill=red,fill opacity=0.5] (\lc,-\firstradius) arc(-90:90:\firstradius) -- (\lc,2*\firstradius)
arc(90:-90:2*\firstradius) -- (\lc,-2*\firstradius);
\draw[very thick] (\lc,-\arc1) arc (-90:90:\arc1);
\node[anchor=west] at (\lc,-\maxradius-0.5) {Iteration 0};
\foreach \s in {1,...,\numradii} {
  \draw[gray] (\rc,-\s*\firstradius) arc(-90:90:\s*\firstradius);
}
\foreach \ang in {0,...,\numrays} {
  \draw[gray] (\rc,0)+(90 - \ang*180/\numrays:\firstradius) -- ++(90 - \ang*180/\numrays:\maxradius);
}
\draw[fill=red,fill opacity=0.5] (\rc,-\firstradius)
arc(-90:-1*180/\numrays:\firstradius) -- ++(-1*180/\numrays:\firstradius)
arc(-1*180/\numrays:-90:2*\firstradius) -- (\rc,-2*\firstradius);
\draw[fill=red,fill opacity=0.5] (\rc,\firstradius)
arc(90:1*180/\numrays:\firstradius) -- ++(1*180/\numrays:\firstradius)
arc(1*180/\numrays:90:2*\firstradius) -- (\rc,2*\firstradius);
\draw[fill=red,fill opacity=0.5] (\rc,0)+(-2*180/\numrays:2*\firstradius)
arc(-2*180/\numrays:2*180/\numrays:2*\firstradius) --
++(2*180/\numrays:\firstradius)
arc(2*180/\numrays:-2*180/\numrays:\maxradius) -- ++(-2*180/\numrays:-\firstradius);
\draw[very thick] (\rc,-\arc1) arc (-90:90:\arc2 and \arc1);
\node[anchor=west] at (\rc,-\maxradius-0.5) {Iteration 1};
\end{tikzpicture}
\caption{Diagram of how trial horizon-find surfaces can move between cells. The
   thick black lines are the trial surfaces. Cells containing the trial surface
   are shaded red. \textit{Left:} Trial surface for fast-flow iteration 0 that
   passes through all cells in the inner ring. \textit{Right:} Trial surface for
   fast-flow iteration 1 where the surface has moved to change which cells are
   intersected.}
\label{fig:horizon-moving}
\end{figure}

However, we must be careful now because this new trial surface is different from
the previous one, and thus may have moved between cells, as shown in figure
\ref{fig:horizon-moving}. Therefore, the horizon interpolator must now check
again if it has received enough data from the cells so that the interpolation
onto the new trial surface can be completed. If not, the horizon interpolator
must wait for data from additional cells. It can sometimes be the case that
enough cells have sent their evolution variables for the first iteration to
finish (left panel of figure \ref{fig:horizon-moving}), but not for the second
iteration (right panel of figure \ref{fig:horizon-moving}); i.e., if only the
inner ring of cells in the figure have sent data before it is time to do
Iteration 1. This implies our asynchronous algorithm must be able to dynamically
receive data from new cells and store them.

There is also a scenario that often happens where a horizon-find will have
finished, but the horizon interpolator will still be receiving evolution
variables from cells that are slower in their evolution but were not needed to
interpolate to any trial surfaces. Therefore, our asynchronous algorithm must be
smart enough to receive these variables from late cells, determine that they are
no longer needed, and dispose of them to avoid a sort of memory leak similar to
the one described in section \ref{sec:io-complications}. This necessitates keeping
ledgers of horizon-find times that are currently being worked on, that are
waiting to be worked on because a previous horizon-find hasn't finished, and
that have already finished.

Another possible scenario that must be taken into account involves the very
first horizon-find of the simulation. Consider the situation where a cell has
advanced far enough in time so that it is able to send its evolution variables
to the horizon interpolator at two different times; an earlier time and a later
time. Because of Statement \ref{statement:a_before_b}, it is possible that the
first message the horizon interpolator receives is the later time. Upon initial
consideration, this doesn't seem to affect things too much. This could be fixed
by having the cell send the time along with the evolution variables, and then
the horizon interpolator could sort its received data by time. However, now
consider the entirely possible scenario that the evolution variables at the
later time for \textit{all} cells arrive at the horizon interpolator before any
of those at the earlier time. The horizon interpolator must somehow know to
\textit{not} start a horizon-find at the later time until the earlier time
finishes, because it should use the result at the earlier time to form its
initial trial surface at the later time.
Making the horizon interpolator aware of the initial time
unfortunately won't work because horizon-find times are determined dynamically.
So it is not known ahead of time when the first horizon-find will occur. Our
solution to this problem is to send to the horizon interpolator the time at
which the previous horizon-find happened along with the time of the current
horizon-find and the evolution variables. This works because a cell knows when
the previous horizon-find happened. For the first horizon-find time, the cell
sends a sentinel value (e.g. NaN) for the previous horizon-find time, which the
horizon interpolator is programmed to recognize as indicating there's no
previous time. This allows the horizon interpolator to act accordingly in the
scenario described in this paragraph and complete the horizon-finds in order.

The effect of this asynchronous implementation is that the process of actually
finding the horizon is largely decoupled from the evolution. Cells simply send
their evolved variables on their grid points to the horizon interpolator and
continue on evolving. In the synchronous implementation, by contrast, cells
would have to wait for a horizon-find to finish before evolving forward again.
Moreover, in the evolution of a binary black hole system, there are two (and
near merger, three) horizon
finds that need to happen, so in the synchronous implementation cells would have
to wait twice as long. However, in the asynchronous implementation there can
just be two horizon interpolators, one for each horizon, and the only extra cost
incurred by the cells is having to send one additional message containing their
evolved variables to a second process.

The drawbacks with the asynchronous implementation are the code complexity
required to account for the numerous different scenarios that can occur when
messages are received and tasks are run in an effectively random order, and the
necessity of retaining copies of the volume data for use by the interpolator
after the cells have evolved past each horizon-find time.

\section{Synchronous control system\label{sec:sync-control-system}}

We next move on to explaining how the horizon finder communicates back to the
black hole evolutions using a feedback control system. We start by explaining
the synchronous aspects of this algorithm as there are a number of them that are
separate from the asynchronous aspects. The synchronous parts are largely the
same as described in refs.~\cite{Scheel:2006gg} and~\cite{Hemberger:2012jz}, and
we summarize them here to point out a few subtle differences that are necessary
for an asynchronous implementation.  We detail the asynchronous parts in
section \ref{sec:async-control-system}.

\spectre~evolves the first-order Generalized Harmonic (FOGH) formulation of
Einstein's equations~\cite{Lindblom:2005qh} using the Damped Harmonic (DH) gauge
condition~\cite{Szilagyi:2009qz, Deppe2018:uye}. In order to handle a
singularity, \spectre~excises a portion of the computational domain around each
singularity. However, this excision introduces a new external boundary of the
computational domain. Normally, this would require that we provide some sort of
boundary condition for this new boundary. However, we place the boundary inside
the apparent horizon of a black hole, and we appropriately control the relative
velocity of this boundary with respect to the apparent horizon so that all
characteristic fields of the system at this boundary are flowing out of the
computational domain and into the black hole. In this case, mathematically a
boundary condition is \textit{not} required there, so we do not impose one.
However, if the position or velocity of the excision boundary changes in a way
such that the characteristic fields are no longer all outgoing, the simulation
will fail because of the missing boundary condition. In order to avoid this, we
employ a feedback control system to dynamically adjust the position, shape, and
velocity of the excision boundary using time-dependent coordinate mappings to
keep the excision within the apparent horizon and keep the characteristic fields
outgoing (into the hole).

\subsection{Computational Domain\label{sec:domain}}
Since the problem we'd like to parallelize is the computation of the FOGH
equations, our domain decomposition is necessarily intimately tied to how we
distribute the work among our computational resources. Asynchronous parallelism
is best utilized with many small tasks that can run quickly. A ``small''
task here
means the work done to compute the time step for a number of grid points in the
computational domain is small compared to the work to take a time step
for all the grid points. The benefit of
asynchronous parallelism is wasted if you have a small number of large tasks
because then you can only utilize a limited number of computational resources;
i.e., the problem won't scale well. Alternatively, if a task is \textit{too}
small, then the overhead of sending/receiving asynchronous messages can dominate
the cost of the simulation. Therefore, a balance must be struck.

To ensure we have many adequately small tasks, \spectre's computational domain
is broken up into partitions called \textit{blocks}. A block can be refined into
numerous cells called \textit{elements}. Each element contains grid points that
are stored contiguously in memory. The job of an element is to then time
integrate the FOGH equations at its grid points. This job is accomplished by
running numerous tasks on the element. An example of a task could be an element
receiving and storing flux information from a neighboring element, then
attempting to integrate the FOGH equations one step further in time. The bulk of
the tasks run on an element should, in some way, attempt to time integrate the
FOGH equations. While an element is executing a task, it can also launch tasks
such as asynchronous I/O \ref{sec:io} or horizon finding
\ref{sec:horizon-finder} on other processors.

\begin{figure}
  \includegraphics[width=0.4\linewidth]{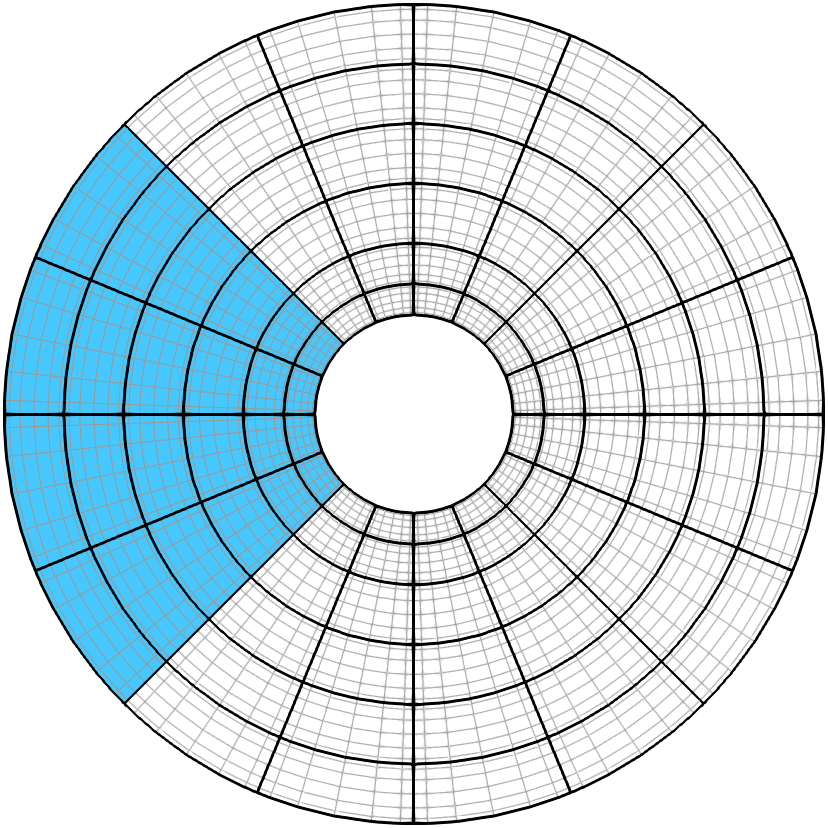}
  \hspace{0.5cm}
  \includegraphics[width=0.4\linewidth]{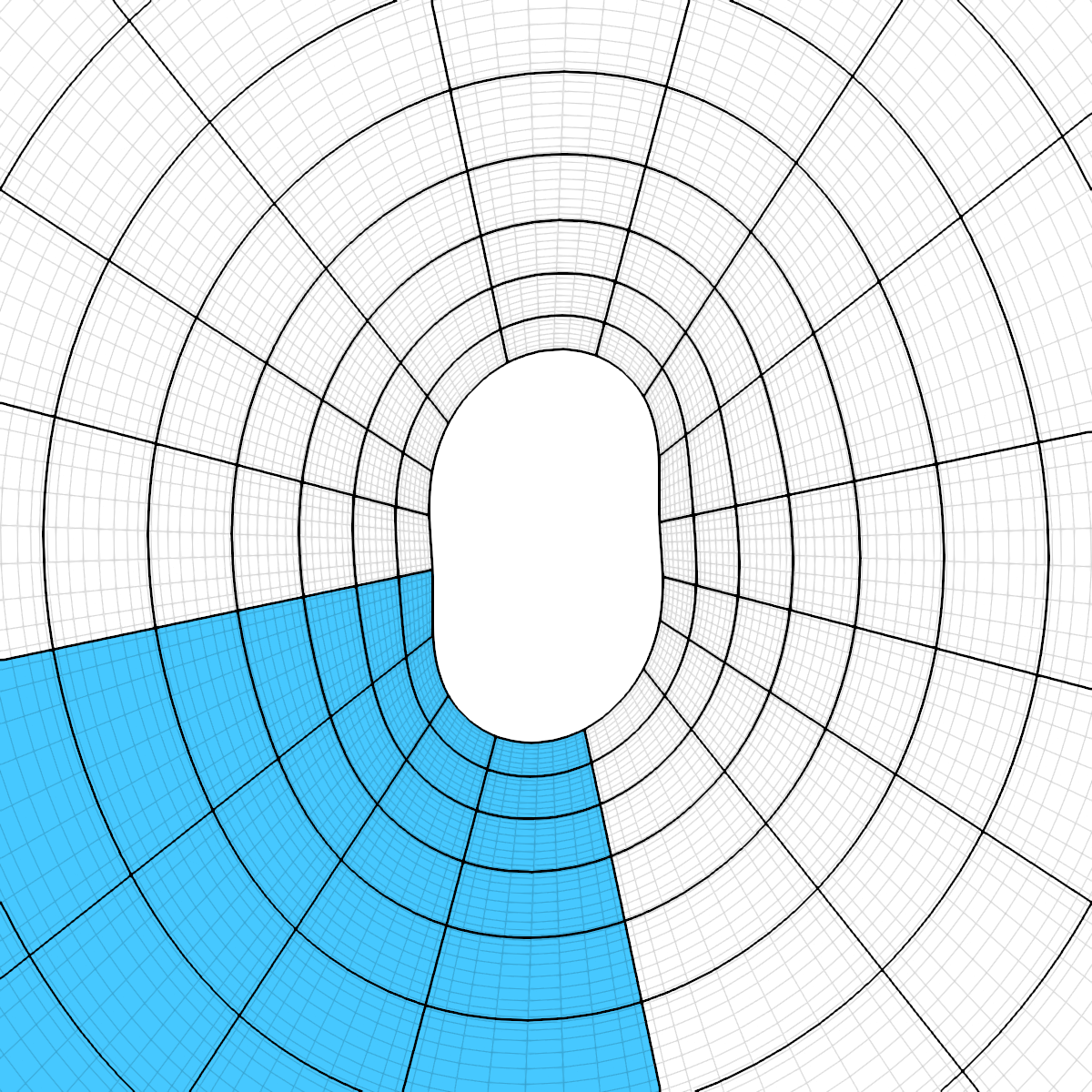}
  \caption{Equatorial slices of 3D excised spherical computational domains used
  for the evolution of single black holes in \spectre\cite{Lovelace:2024wra}.
  The thicker black lines represent element boundaries, and the thinner grey
  lines represent grid points. The blue shaded regions represent blocks of
  \spectre's domain. \textit{Left:} An undistorted spherical domain.
  \textit{Right:} A spherical domain with time dependent coordinate mappings
  distorting and rotating the excision.}
  \label{fig:sphere-highlighted}
\end{figure}

Figure \ref{fig:sphere-highlighted} shows examples of computational domains with
excisions in \spectre~used for single black hole evolutions. The thick black
lines are element boundaries and the thin grey lines are grid points. The blue
shaded regions are the equatorial projections of a single block. The coordinates
for these grid points in figure \ref{fig:sphere-highlighted} are considered the
grid coordinates, defined in section \ref{sec:time-dep-map}. The central
excision surface must be kept within an apparent horizon for the entire
evolution.

\subsection{Time-dependent coordinate mapping\label{sec:time-dep-map}}

In order to keep an excision surface inside an apparent horizon throughout an
evolution, we implement time-dependent coordinate mappings that move and distort
the computational domain. An example of a domain after time-dependent coordinate
mappings have been applied is shown in the right panel of figure
\ref{fig:sphere-highlighted}. Our implementation closely follows that of
\cite{Scheel:2006gg} and \cite{Hemberger:2012jz} except for a few subtle (yet
critical) differences to allow for use with asynchronous parallelism. Here we
give enough of a summary of refs.~\cite{Scheel:2006gg}
and~\cite{Hemberger:2012jz} so that we can explain the necessary changes.

We start with a computational mesh that is described by coordinates $\{\hat{t},
\xi^{i}\}$ which we will call the ``grid coordinates''. These grid coordinates
are stationary and do not change over the course of an evolution. They can be
thought of as a sort of comoving coordinates. An example of grid coordinates for
a single black hole is shown in the left panel of
figure~\ref{fig:sphere-highlighted}. We then apply a coordinate mapping of the
form
\begin{eqnarray}
  t &=& \hat{t} \\
  x^{i} &=& \mathcal{M}(\hat{t}) \xi^{i}
\end{eqnarray}
where $\mathcal{M}(\hat{t})$ is a (generally) non-linear operator acting on the
grid coordinates.
We call $\{t, x^{i}\}$ the ``inertial coordinates'' as they represent
coordinates where a binary system would orbit and have the separation shrink. An
example of inertial coordinates for a single black hole is shown in the right
panel of figure~\ref{fig:sphere-highlighted}. From now on, we will only use $t$,
not $\hat{t}$, when referring to the coordinate time since they are the same.

The spatial mapping $\mathcal{M}(t)$ is a composition of analytic functions
applied to the grid coordinates $\xi^{i}$. For example, a 2D rotational mapping
$\mathcal{M}_{R2D}$ could take the form
\begin{equation}
  \mathcal{M}_{R2D} :
  \begin{pmatrix}
    x^{1} \\
    x^{2}
  \end{pmatrix}
  =
  \begin{pmatrix}
   \cos\theta(t) & -\sin\theta(t) \\
   \sin\theta(t) &~~\cos\theta(t)
  \end{pmatrix}
  \begin{pmatrix}
    \xi^{1} \\
    \xi^{2}
  \end{pmatrix},
\end{equation}
where $\theta(t)$ is a time-dependent function that evaluates to the rotation
angle at coordinate time $t$. Each block contains its own mapping
$\mathcal{M}_{B}(t)$, with the only requirement that neighboring blocks'
mappings be continuous at the boundaries. They need not be differentiable at
block boundaries. Certain mappings can cover the entire computational domain,
i.e.~a global mapping; or they can cover only part of the computational domain,
i.e.~a quasi-local mapping. We use the term ``quasi-local'' and not ``local''
because each element shares the same mapping as its parent block. Therefore,
even if a mapping is in only one block of the computational domain, it is likely
shared across multiple elements, and therefore across computational resources.
Also, while a mapping may only be quasi-local and not completely global in the
topological sense, it can be thought of as global in the computational sense
since it can span multiple processors and often times multiple nodes.

We can generalize this to any mapping $\mathcal{M}(t) \equiv
\mathcal{M}(\lambda(t))$ where $\lambda(t)$ is a general time-dependent
function. We cannot choose $\lambda(t)$ to be some known analytic function
because we do not know the trajectory or shape of the apparent horizon for the
entire inspiral and merger. The trajectory and shape of the apparent horizons
are determined by the solution of Einstein's equations we are trying to obtain.
Even post-Newtonian approximations~\cite{Blanchet:2024mnz} for the trajectories
are not accurate enough for our needs since the error in the trajectory would
accrue throughout the inspiral.

Instead we choose to represent $\lambda(t)$
as a piecewise polynomial of the form
\begin{equation}
  \label{eq:lambda}
  \lambda(t) = \sum_{n=0}^{N}\frac{1}{n!}(t-t_{i})^n\lambda_{i}^{(n)}
    \quad\quad\forall\quad t_{i} < t \leq t_{i+1}
\end{equation}
with $N$ being the degree of each piece of the polynomial and
$\lambda_{i}^{(n)}$ are constants within $t_{i} < t \leq t_{i+1}$. The reason
for this choice of functional form for $\lambda(t)$ is explained in section
\ref{sec:control-error-signal}. Note that we put $(n)$ in parentheses in
$\lambda_{i}^{(n)}$ to signify that it is not an exponent, but just a label. At
$t_{i+1}$ we choose a new set of constants $\lambda_{i+1}$ such that
\begin{eqnarray}
  \label{eq:lambda-i-N}
  \lambda_{i+1}^{(N)} &=& \frac{d^{N}\lambda}{dt^{N}} = U(t_{i+1}) \\
  \label{eq:lambda-i-n}
  \lambda_{i+1}^{(n)} &=&
    \frac{d^{n}\lambda}{dt^{n}}\biggm|_{t=t_{i+1}}\quad\quad\forall\quad0 \leq n
    < N
\end{eqnarray}
where $U(t_{i+1})$ will be defined in section \ref{sec:control-error-signal}.
This ensures that all derivatives of $\lambda$ are continuous except the $N$th
derivative.

There is now an ambiguity of which set of coefficients to use at time $t_{i+1}$,
i.e.~$\lambda_i^{(n)}$ vs $\lambda_{i+1}^{(n)}$. We choose to use the
coefficients $\lambda_i^{(n)}$ at $t_{i+1}$ to ensure a consistent evolution.
Note that during the simulation, we write the values of all $t_{i}$ and
$\lambda_i^{(n)}$ to disk, so that any postprocessing application that uses the
output of the simulation will be able to reconstruct the inertial coordinates at
any arbitrary time. When postprocessing a simulation, we want the coordinate
mappings to return the exact same values at a given time as they did during the
evolution. If we used the $\lambda_i^{(n)}$ coefficients at $t_{i+1}$ during the
evolution (because the new ones hadn't been computed yet), but during the
postprocessing we used the $\lambda_{i+1}^{(n)}$ coefficients (because we now
had them), this would produce different values for the coordinate mappings at
the same time. Our postprocessed data would then be inconsistent with the data
from the evolution. Additionally, choosing $\lambda_{i+1}^{(n)}$ is necessary
for an asynchronous implementation (see section \ref{sec:async-control-system}).

\subsection{Control error and signal\label{sec:control-error-signal}}

For the time-dependent coordinate mappings $\mathcal{M}(\lambda(t))$ described
in section \ref{sec:time-dep-map}, we choose the form of $\lambda(t)$ in
Eq.~\eqref{eq:lambda} because we \textit{a priori} cannot predict the motion and
shape of the excision required to stay inside the apparent horizon. Therefore,
$\lambda(t)$ needs to be dynamically updated as the simulation progresses based
on the evolution of the apparent horizons. We achieve this by applying control
theory to the time-dependent coordinate mappings in order to dynamically update
the $N$th derivative of $\lambda(t)$.

We follow the definitions described \S 3.1 of \cite{Hemberger:2012jz}. Define
a control error $Q$ to be a measure of how incorrect a map parameter
$\lambda(t)$ is.
If we can define a target value $\lambda_{\textrm{target}}$ that
does not depend on $\lambda(t)$, then we define
\begin{equation}
  \label{eq:control-error-linear}
  Q = \lambda_{\textrm{target}} - \lambda(t).
\end{equation}
If we cannot define $\lambda_{\textrm{target}}$ because our system is nonlinear
and the target value would depend on $\lambda(t)$, we require that
\begin{equation}
  \label{eq:control-error-nonlinear}
  \frac{\partial Q}{\partial \lambda} = -1 + \mathcal{O}(Q)
\end{equation}
In both cases we want to drive $Q\rightarrow 0$ so that $\lambda(t)$ (i.e.~the
position and shape of the excision boundary) is driven towards
$\lambda_\textrm{target}$ (i.e.~position and shape of the apparent horizons).

Next, we define the control signal $U(t)$ which we will use to reset the highest
derivative of the polynomial representation of $\lambda(t)$ in
Eq.~\eqref{eq:lambda-i-N}. We use a proportional $N$th derivative (PND)
controller of the form
\begin{equation}
  \label{eq:control-signal}
  U(t) = \sum_{k=0}^{K} a_{k}\frac{d^{k}Q}{dt^{k}},
\end{equation}
where $a_{k}$ are constants that are chosen to critically damp $Q$ to 0 on a
timescale $\tau_{d}$. Typically $K = 2$. The reason we use a PND controller
instead of the typical proportional-integral-derivative (PID) controller used in
numerous other applications of control theory is from years of experience
using a PND controller in \spec~\cite{Hemberger:2012jz}.

In order for this scheme to remain stable, we
demand that $\tau_{d} > t_{i+1} - t_{i}$. This allows us to assume
that $Q(t)$ and $\lambda(t)$ are approximately
smooth over the timescale $\tau_{d}$.
With this assumption we are now able to compute the time derivatives of $Q(t)$
and the coefficients $a_{k}$ necessary for computing $U(t)$ in
Eq.~\eqref{eq:control-signal}.
It is shown in ref.~\cite{Hemberger:2012jz} that $U(t)$ is given by
\begin{equation}
  \label{eq:UofTdef}
  U(t) = \frac{d^{N}\lambda}{dt^{N}} = -\frac{d^{N}Q}{dt^{N}},
\end{equation}
which can be substituted into Eq.~\eqref{eq:control-signal} and then solved to get
$Q\propto e^{-t/\tau_d}$. To get a critically damped solution for a typical $K
= 2, N = 3$ PND system, we choose $a_{0} = 1/\tau_{d}^{3}$, $a_{1} =
3/\tau_{d}^{2}$, and $a_{2} = 3/\tau_{d}$.

In order to compute the time derivatives of $Q(t)$, we measure and record $Q(t)$
for $M$ times between each $t_i$ and then use Lagrange interpolation polynomials
to get $d^{k}Q/dt^{k}$. Typically we use $M = N + 1$ and use $K\textrm{th}$
order non-uniformly spaced stencils for the derivatives. For parameters that
have small $\tau_d$, the measurements are often done once per time step. For
parameters with larger $\tau_d$, measurements can be done less frequently.
Since $\tau_d$ can dynamically change over the course of a simulation,
how often we measure $Q(t)$ will change as well.

There are additional procedures outlined in \S 3 of \cite{Hemberger:2012jz}
related to averaging out noisy measurements of $Q(t)$ and also dynamically
adjusting the damping timescale $\tau_{d}$ of a control system. These are also
implemented in \spectre, but neither of these affect the aspects of the control
system and horizon finding related to synchronous parallelism. Accordingly, we
refer the reader to \cite{Hemberger:2012jz} for details. Dynamically adjusting
the damping timescales can affect the asynchronous parallelism and will be
discussed in section \ref{sec:async-control-system}.

\subsection{Computing $\lambda^\mathrm{new}(t)$\label{sec:sync-update-lambda}}

Now that we have gone through the necessary pieces of horizon finding (section
\ref{sec:horizon-finder}), the computational domain (section \ref{sec:domain}),
the time-dependent coordinate mappings (section \ref{sec:time-dep-map}), and the
control signal (section \ref{sec:control-error-signal}), we detail the process
that occurs in a synchronous application to dynamically update the
time-dependent coordinate mappings. For simplicity, we will assume the
following: there is only one time-dependent map parameter $\lambda(t)$ that we
are controlling; the order of $\lambda(t)$ is $N = 3$; the order of the PND
controller is $K = 2$; we measure $Q(t)$ $M = 4$ times; and $\tau_d$ is small
enough that we measure $Q(t)$ every time step.

In the previous section, we discuss and treat $\lambda(t)$ as a function that
spans the entire evolution that we periodically update. We will continue to do
so, but we also define two new quantities, $\lambda^\mathrm{new}(t)$ and
$\lambda^\mathrm{old}(t)$, for convenience. $\lambda^\mathrm{new}(t)$ refers to
$\lambda(t)$ once the new coefficients $\lambda_{i+1}$ have been computed and
used. $\lambda^\mathrm{old}(t)$, then refers to $\lambda(t)$ before these new
coefficients have been computed.

The (synchronous) steps to compute $\lambda^\mathrm{new}(t)$ are:

\begin{enumerate}
  \item \label{lst:lambda-1} Wait for all elements to finish their time step.
  Evolution variables are now in a consistent state at the same time.
  \item Find the apparent horizon (section \ref{sec:horizon-finder}).
  \item Calculate and store $Q(t)$ using the apparent horizon
  \item If not the $M=4\textrm{th}$ measurement, go to step \ref{lst:lambda-1}
  \item At the $M=4\textrm{th}$ measurement, compute $U(t)$ using
  $d^{k}Q/dt^{k}$ and compute a new $\tau_d$.
  \item Use $U(t)$ to set $d^{N}\lambda/dt^{N}$ at $t_{i+1}$.
  \item Use $\tau^\mathrm{new}_d$ to possibly set the new time step because we
  require that $\tau_d > t_{i+1} - t_i$. Broadcast $d^{N}\lambda/dt^{N}$ and new
  time step to all nodes/cores.
  \item On each node that receives $d^{N}\lambda/dt^{N}$, use Eqs.
  \eqref{eq:lambda-i-N}-\eqref{eq:lambda-i-n} and $\lambda^\mathrm{old}(t)$
  to compute $\lambda^\mathrm{new}(t)$.
  \item Go to step \ref{lst:lambda-1}
\end{enumerate}

With this procedure, all elements must wait until the horizon-find finishes and
until all control system quantities are computed and broadcast before continuing
the evolution. The advantage of the synchronous implementation is that
everything happens deterministically and in order, making the logic-flow of the
code easy to follow. The disadvantage is that during the $M=1,2,3$ measurements
of $Q(t)$, which include horizon-finds, no global communication of
time-dependent parameters needs to happen. This implies the elements could, in
theory, keep evolving until the $M=4$ measurement without stopping and waiting
(assuming the horizon finder is implemented asynchronously as in section
\ref{sec:horizon-finder}). This is precisely the performance improvement we
employ with our asynchronous approach in section \ref{sec:async-control-system}.

\section{Asynchronous control system\label{sec:async-control-system}}

Here we detail the differences from the previous section
when implementing an asynchronous feedback control
system that mutates a global quantity in a simulation; i.e., dynamically
updating the time-dependent map parameters in \spectre.

\subsection{Global State\label{sec:global-state}}

In section \ref{sec:time-dep-map}, we briefly mentioned that the time-dependent
coordinate mappings can be quasi-local in the sense that they can cover multiple
elements. In turn, because elements are then placed on many cores/nodes, this
means the mappings can also span multiple cores/nodes. Therefore the mappings
need to be kept consistent across the computational resources in order to have a
consistent evolution. This implies that the time-dependent coordinate mappings
(or more accurately the time-dependent map parameters $\lambda(t)$) are what we
call a ``global state'' of the simulation. Such a global state
wouldn't be an issue if it were constant or immutable, because then it could just
be set at the beginning and would never change during the simulation.
However, $\lambda(t)$ \textit{does} need to be mutated
using the feedback control system. Therefore, we don't just have
a global state, but a mutable global state. A mutable global state can cause
issues in an asynchronous application because keeping the state of the simulation
consistent can be quite difficult and error prone.

One of the common wisdoms when dealing with any asynchronous application is to
avoid a mutable global state whenever possible. However, in our case we cannot
avoid such a global state because of the nature of our quasi-local
time-dependent coordinate mappings. One common way of dealing with a mutable
global state when porting a synchronous application to be asynchronous is called
privatization of the global state \cite{Solihin2020-en}. Instead of the global
state being located in a common place accessible to all tasks, a local copy of
the global state is stored in each task. Then, a task simply continues until
it's local copy of the global state is invalid and/or it receives a new global
state.

While this is theoretically possible in our case, privatization becomes
impractical with a large number of tasks and with a global state that grows in
size over time. Our production-level simulations will typically have $10^3$ --
$10^5$ elements (tasks) and the size of $\lambda(t)$ (as parameterized by the
coefficients $\lambda_i^{(n)}$) will grow to around $75\,$MB by the end of the
simulations (because of how many times we mutate $\lambda(t)$ and the necessity
to store all past history described at the end of section
\ref{sec:time-dep-map}). If we were to store $\lambda(t)$ locally on each task,
this would require $75\,$GB -- $7.5\,$TB of total memory. Even for modern
high-performance computing systems, this is likely too much. Therefore we keep
only a single copy of $\lambda(t)$ in a shared place on a node, mutate that
copy, and algorithmically ensure access to this shared mutable global state is
consistent. An alternative solution which we do not employ would be to use
privatization of $\lambda(t)$ but only store recent history in memory, and
periodically write older history to disk and purge it from memory.

\subsection{Expiration time\label{sec:exp-time}}

Before we detail how we mutate $\lambda(t)$ consistently, we must explain one of
the most critical (but also subtle) differences between the synchronous and
asynchronous control system implementations, which comes from $t_{i+1}$ in
Eq.~\eqref{eq:lambda}. In the synchronous case,
because the evolution was paused at
$t_{i+1}$ in order to find the horizon and then compute
$\lambda^\mathrm{new}(t)$, no elements will evolve past $t_{i+1}$ and try to evaluate
$\lambda(t)$ at a time $t > t_{i+1}$. Consistency in using $\lambda(t)$ is thus
built into the synchronous algorithm.

However, in the asynchronous case this is now an issue. Finding the apparent
horizon is an asynchronous algorithm, so the elements won't pause to wait for a
horizon-find to finish. Therefore, so long as they are able, the elements will
keep evolving forward in time and eventually past $t_{i+1}$. This can cause an
inconsistent state in the simulation. Take the case where an element will evolve
past $t_{i+1}$, the $M=4$ measurement of the horizon occurs, and
$\lambda^\mathrm{new}(t)$ is computed and broadcast. If
$\lambda^\mathrm{new}(t)$ arrives at the node the element is on \textit{before}
the element evolves past $t_{i+1}$, then the element will use
$\lambda^\mathrm{new}(t)$ for its evolution. However, if the
$\lambda^\mathrm{new}(t)$ arrives \textit{after} the element evolves past
$t_{i+1}$, then the element will have used the outdated
$\lambda^\mathrm{old}(t)$ for its evolution and will be incorrect. Furthermore,
with many elements in a simulation on different nodes, it's possible that some
elements use the correct $\lambda^\mathrm{new}(t)$ and some use the incorrect
$\lambda^\mathrm{old}(t)$ because of Statement \ref{statement:a_before_b} and
message ordering, resulting in a completely inconsistent evolution.

To address this potential inconsistency, we conceptually promote $t_{i+1}$ from
being the end of the interval of $\lambda(t)$ to what we call an ``expiration
time''. With this promotion, we require that elements cannot evolve past an
expiration time. Instead, if they otherwise would be able to evolve past the
expiration time, they must pause and wait to receive $\lambda^\mathrm{new}(t)$
before continuing. In practice we have many different $\lambda(t)$ for the
various time-dependent coordinate mappings \spectre~uses, so an element must
obey the expiration time of each $\lambda(t)$ (i.e., the $\min\{t^j_{i+1}\}$).
In computer science language, we use $t_{i+1}$ to establish a
\textit{happens-before} relation between tasks.

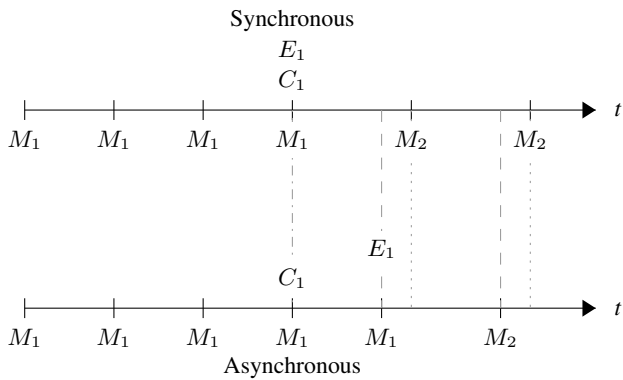
\begin{figure}
\begin{tikzpicture}[x=0.75pt,y=0.75pt,yscale=-1,xscale=0.75]
\tikzmath{
  \tickh = 10;
  \labelh = 15;
  \syncymid = 108;
  \syncybottom = \syncymid - \tickh/2;
  \syncytop = \syncymid + \tickh/2;
  \syncyM = \syncymid + \labelh;
  \syncyC = \syncymid - \labelh;
  \syncyE = \syncymid - 2*\labelh;
  \syncytitle = \syncymid - 3*\labelh;
  \asyncymid = 208;
  \asyncybottom = \asyncymid - \tickh/2;
  \asyncytop = \asyncymid + \tickh/2;
  \asyncyM = \asyncymid + \labelh;
  \asyncyC = \asyncymid - \labelh;
  \asyncyE = \asyncymid - 2*\labelh;
  \asyncytitle = \asyncymid + 2*\labelh;
  \deltax1 = 60;
  \deltax2 = 80;
  \tickAx = 128;
  \tickBx = \tickAx + \deltax1;
  \tickCx = \tickBx + \deltax1;
  \tickDx = \tickCx + \deltax1;
  \tickEsyncx = \tickDx + \deltax2;
  \tickFsyncx = \tickEsyncx + \deltax2;
  \tickEasyncx = \tickDx + \deltax1;
  \tickFasyncx = \tickEasyncx + \deltax2;
  \endx = 512;
  \tx = \endx + \labelh;
}

\draw (\tickAx, \syncymid) -- (\endx,\syncymid);
\foreach \x in {\tickAx, \tickBx, \tickCx, \tickDx, \tickEsyncx, \tickFsyncx} {
  \draw (\x,\syncybottom) -- (\x,\syncytop);
}
\draw [shift={(\endx,\syncymid)}, rotate = 180] [fill=black][line width=0.08]
[draw opacity=0] (8.93,-4.29) -- (0,0) -- (8.93,4.29) -- cycle;

\draw (\tickAx, \asyncymid) -- (\endx,\asyncymid);
\foreach \x in {\tickAx, \tickBx, \tickCx, \tickDx, \tickEasyncx, \tickFasyncx} {
  \draw (\x,\asyncybottom) -- (\x,\asyncytop);
}
\draw [shift={(\endx,\asyncymid)}, rotate = 180] [fill=black  ][line width=0.08]
[draw opacity=0] (8.93,-4.29) -- (0,0) -- (8.93,4.29) -- cycle;

\tikzset{
longvert line/.style={color={rgb, 255:red, 155; green, 155; blue, 155 },draw
opacity=1,fill={rgb, 255:red, 0; green, 0; blue, 0 }  ,fill opacity=0},
dashed/.style={longvert line, dash pattern={on 4.5pt off 4.5pt}},
dotted/.style={longvert line, dash pattern={on 0.84pt off 2.51pt}},
dashdot/.style={longvert line, dash pattern={on 0.75pt off 2.25pt on 3.75pt off
2.25pt}}
}
\draw [dashdot] (\tickDx,\syncymid) -- (\tickDx,\asyncymid);
\draw [dashed] (\tickEasyncx,\syncymid) -- (\tickEasyncx,\asyncymid);
\draw [dotted] (\tickEsyncx,\syncymid) -- (\tickEsyncx,\asyncymid);
\draw [dashed] (\tickFasyncx,\syncymid) -- (\tickFasyncx,\asyncymid);
\draw [dotted] (\tickFsyncx,\syncymid) -- (\tickFsyncx,\asyncymid);

\foreach \x in {\tickAx, \tickBx, \tickCx, \tickDx} {
\draw (\x,\syncyM) node [align=center,fill=white] {$M_1$}; }
\foreach \x in {\tickEsyncx, \tickFsyncx} {
\draw (\x,\syncyM) node [align=center,fill=white] {$M_2$}; }
\draw (\tickDx,\syncyC) node [align=center] {$C_1$};
\draw (\tickDx,\syncyE) node [align=center] {$E_1$};
\draw (\tx,\syncymid) node [align=left] {{\fontfamily{ptm}\selectfont
\textit{t}}};

\foreach \x in {\tickAx, \tickBx, \tickCx, \tickDx, \tickEasyncx}
{ \draw (\x,\asyncyM) node [align=center,fill=white] {$M_1$}; }
\draw (\tickFasyncx,\asyncyM) node [align=center,fill=white] {$M_2$};
\draw (\tickDx,\asyncyC) node [align=center,fill=white] {$C_1$};
\draw (\tickEasyncx,\asyncyE) node [align=center,fill=white] {$E_1$};
\draw (\tx,\asyncymid) node [align=left] {{\fontfamily{ptm}\selectfont
\textit{t}}};

\draw (\tickDx,\syncytitle) node [align=left] {{\fontfamily{ptm}\selectfont
Synchronous}};
\draw (\tickDx,\asyncytitle) node [align=left] {{\fontfamily{ptm}\selectfont
Asynchronous}};
\end{tikzpicture}
\caption{Timeline for computing $\lambda^\mathrm{new}(t)$ for both the
synchronous (\textit{top}) and asynchronous (\textit{bottom}) case. Here $M_j$
denotes when we perform a measurement (horizon-find), $C_j$ denotes when we
compute $\lambda^\mathrm{new}(t)$, and $E_j$ denotes the expiration time which
is when we mutate $\lambda^\mathrm{old}(t)\rightarrow\lambda^\mathrm{new}(t)$
and set a new expiration time. In the synchronous case, we compute and mutate
$\lambda(t)$ at the same time because all elements are already paused. In the
asynchronous case, we delay mutating $\lambda(t)$ by one (old) measurement for
better utilization. The vertical dashed and dotted lines show how measurements
(fail to) line up between the synchronous vs.\ asynchronous cases.}
\label{fig:exp-time}
\end{figure}

Even though we have fixed the consistency issue, we need to also consider
efficiency and utilization when obeying the expiration time. For the synchronous
case, we perform a horizon-find at the same time as we compute and mutate
$\lambda^\mathrm{new}(t)$. Again, for the synchronous case this is irrelevant
because the evolution of the elements is already paused. However, when we
consider the asynchronous case, if we keep the same structure of requiring a
horizon-find at the expiration time, the elements will send their data to the
asynchronous horizon finder and then immediately have to wait for
$\lambda^\mathrm{new}(t)$. This is not efficient, since a horizon-find will
always have to happen at the expiration time and elements will always have to
wait, causing low utilization of the computational resources. A different
mutation structure is needed in order to minimize how much time the elements
wait.

This new mutation structure is depicted in figure \ref{fig:exp-time}. The
horizontal axes represent time. The top axis is for the synchronous case, and
the bottom axis is for the asynchronous case. Here $M_j$ denotes when we perform
a measurement (horizon-find), $C_j$ denotes when we compute
$\lambda^\mathrm{new}(t)$, and $E_j$ denotes the expiration time $t_{i+1}$ when
we actually mutate $\lambda^\mathrm{old}(t)\rightarrow\lambda^\mathrm{new}(t)$.
For the synchronous case, $C=E=t_{i+1}$ and the first four measurements are
equally spaced at $\Delta t_{M_1}$ while the subsequent two are also equally
spaced but at $\Delta t_{M_2} > \Delta t_{M_1}$ to denote a change in $\tau_d$.

For the asynchronous case, when we start a measurement that will eventually
compute $\lambda^\mathrm{new}(t)$, we don't want the elements to stop
immediately. We'd like to have them continue evolving for a short while so the
horizon-find has a chance to finish and we can mutate
$\lambda^\mathrm{old}(t)\rightarrow\lambda^\mathrm{new}(t)$ before the element
reaches the expiration time. Therefore we need to decouple $C$ and $E$ in
figure~\ref{fig:exp-time} for the asynchronous case.
We achieve this by delaying $E$ by
one extra measurement of $\lambda^\mathrm{old}(t)$. Then, when we start the
measurement $C$ that will be used to compute $\lambda^\mathrm{new}(t)$, the
elements still have time left before they hit the expiration time.
They can evolve, allowing forward evolution progress to happen while
the horizon-find for $C$ happens.

The choice of how long to delay $E$ is somewhat arbitrary. If it's too far into
the future, outdated horizon information will be used to compute
$\lambda^\mathrm{new}(t)$; too soon after $C$ and the elements will be quickly
blocked by the expiration time. We found one (old) measurement time to be a
suitable balance between these two concerns. Additionally, this choice fits
nicely into the existing implementation infrastructure.

We note that this change of delaying the mutation by one (old)
measurement time means that we will be using ``outdated'' information when we
mutate $\lambda(t)$ at $E$. In the bottom axis of figure \ref{fig:exp-time} this
is shown with the measurement $M_1$ which occurs at the same time as $E_1$ but
will be used as the first measurement for computing $C_2$ (not shown). This
measurement $M_1$ is performed with $\lambda^\mathrm{old}(t)$, not
$\lambda^\mathrm{new}(t)$, and therefore uses ``outdated'' information. An
improvement could be to extrapolate the value computed at $C$ to the time of
$E$. Then, this extrapolated value $C^\mathrm{extrap}$ will act as if we used
the measurement $M$ at $E$ to compute $C$ (i.e.~them all occurring at the same
time like the synchronous case) while still maintaining the asynchronous
advantage of delaying the mutation of $\lambda(t)$ by one (old) measurement.

\subsection{Querying $\lambda(t)$ and computing $\lambda^\mathrm{new}(t)$\label{sec:async-update-lambda}}

Now that we have an efficient prescription for when to compute
$\lambda^\mathrm{new}(t)$ and mutate it, the last piece we need is a method to
actually query $\lambda(t)$ and then perform the mutate in a consistent way. As
a reminder, $\lambda(t)$ is a shared mutable global state, meaning that many
different elements will all access the same copy of $\lambda(t)$ in memory on a
node. Additionally, all $\lambda(t)$ are stored in a single data structure we'll
denote as $\Lambda(t) = \{ \lambda_k(t) \}$. This is because we actually have
several time-dependent coordinate mappings, each using one or multiple of the
$\lambda_k(t)$. Therefore, care needs to be taken when mutating $\Lambda(t)$
since elements may be using it simultaneously.

The most naive method for ensuring this consistency is to allow only one
element/task to query or mutate $\lambda(t)$ (or $\Lambda(t)$) at a time.
However, this would be extremely inefficient, especially for a large number of
elements on a single node. Therefore, we put two restrictions on how
$\lambda(t)$ can be queried and mutated, which will inform what data structure
we can use to represent $\Lambda(t)$. We first require that $\Lambda(t)$ cannot
be mutated by two different cores simultaneously (i.e.~two different
$\lambda_k(t)$). While not strictly necessary, this assumption simplifies the
code significantly and we have not found this to affect performance. The second
requirement is that each $\lambda_k(t)$ must be queryable at all wallclock
times, even while it is being mutated. A useful case is for elements that are
slower than others. Slower elements may need to query one $\lambda_k(t)$ (and
thus don't depend on a mutate to $\Lambda(t)$) and should be able to do so
without waiting for a mutation of a different $\lambda_l(t)$ to finish.

A data structure that satisfies the above requirements is a linked list for each
$\lambda_k(t)$ where access to the most recent entry of each list is protected
by an atomic variable. Then those lists are stored in a map data structure.
Since we store all past history, these linked lists will only be appended to and
their entries will never be changed. The thread safety of each list when being
queried and mutated simultaneously is guaranteed by the atomic variable, which
is more efficient than locks.

\LinesNumbered
\begin{algorithm}
\SetAlgoNoLine
\DontPrintSemicolon
\caption{Registering a callback with $\Lambda(t)$}\label{alg:lambda}
  \eIf{$t \leq t_{i+1}$} {
    No callback is registered; value is fetched
  }{\label{alg:lambda-else-1}
    Element registers a callback with $\Lambda(t)$\label{alg:lambda-else-1-body}

    \eIf{$t \leq t_{i+1}$} {\label{alg:lambda-if-2}
      If a callback exists, remove it; value is fetched
    }{
      A callback was already registered. Don't register another
    }
  }
\end{algorithm}

Next, we discuss the procedure for how elements can query $\Lambda(t)$, which
has some subtleties. A key feature for this procedure is the concept of a
``callback''. A callback is some function to be run after some other event
(function, communication, task) has finished its execution. Callbacks are used
in our algorithm when an element checks if it can evaluate $\Lambda(t)$ at a
time after the expiration time. When this happens, the element pauses its
evolution and ``registers'' a callback with $\Lambda(t)$. Registration of a
callback involves locking the list of existing callbacks for $\Lambda(t)$,
appending the incoming callback to the list, then releasing the lock. Once a
$\lambda_k(t)$ is mutated, this callback will restart the evolution of the
element.

The subtleties arise based on the order of events happening; i.e., whether
$\lambda_k(t)$ is being mutated just before or just after the element registers
a callback with $\Lambda(t)$. The details are described with pseudocode in
Algorithm \ref{alg:lambda}. When an element queries $\Lambda(t)$, it first
checks if the time it's querying is after the expiration time. If so, it
registers a callback. If not, no callback is registered. Then we check the same
condition a second time immediately afterwards. This is because of how we define
$\Lambda(t)$. Therefore, suppose that the if-else starting at line
\ref{alg:lambda-if-2} in Algorithm \ref{alg:lambda} was omitted.  Then the
following events could occur sequentially in wallclock time:
\begin{enumerate}
  \item An element queries $\Lambda(t)$ for $\lambda_0(t)$ and registers a
  callback.
  \item $\lambda_1(t)$ is mutated, copies and invokes all callbacks of
  $\Lambda(t)$ restarting the element.
  \item\label{lst:callback-3} The element again queries $\Lambda(t)$ for
  $\lambda_0(t)$ and finds it still hasn't been mutated, but has not yet
  registered a callback (between lines \ref{alg:lambda-else-1} and
  \ref{alg:lambda-else-1-body} in Algorithm \ref{alg:lambda}).
  \item\label{lst:callback-4} $\lambda_0(t)$ is mutated \textit{after} the
  element queried for $\lambda_0(t)$, found no callbacks, and does nothing.
  \item\label{lst:callback-5} Now the callback from the query in step
  \ref{lst:callback-3} is registered, and the element is paused.
\end{enumerate}
We now have a scenario where the element is paused waiting for a change that has
already occurred in step \ref{lst:callback-4}, and our simulation will deadlock
(section \ref{sec:deadlock}). This is why we must check the same condition a
second time after step \ref{lst:callback-5}. This second check is line
\ref{alg:lambda-if-2} in Algorithm \ref{alg:lambda}. When we do this, we'll
conclude that $\lambda_0(t)$ is in fact ready, remove the callback which is now
unneeded, and the element can proceed. If $\lambda_0(t)$ still wasn't ready,
then we have already registered a callback so the element will be paused and we
can be assured that the mutation of $\lambda_0(t)$ will come sometime in future
wallclock time.

Finally, we discuss how to mutate $\Lambda(t)$ which is considerably more
straightforward than querying $\Lambda(t)$ because we only allow one mutation of
$\Lambda(t)$ at a time. To mutate $\Lambda(t)$, we lock $\Lambda(t)$ to prevent
other mutations and append the new $\lambda_i^{(n)}$ coefficients
(Eqs.~\eqref{eq:lambda} -- \eqref{eq:lambda-i-n}) to the linked list of the
$\lambda_k(t)$ that's being mutated. After we release the mutation lock, we lock
the vector of callbacks associated with $\Lambda(t)$, copy them, release the
callback lock, and invoke each callback in the copied list. The reason for the
copy is because some callbacks may immediately register another callback (as
explained in the previous paragraph), so we need to avoid an infinite recursion
of callbacks being invoked and then those same callbacks being re-registered.

Additionally, if a callback is registered between the wallclock time when we
release the mutation and acquire the callback lock, this is not an issue.
Whether $\Lambda(t)$ is ready or not, this newly registered callback will still
be called. If $\Lambda(t)$ is ready, then the callback will likely restart the
evolution of an element. If a $\lambda_k(t)$ still isn't ready, then a new
callback will just be registered for when a different $\lambda_l(t)$ in
$\Lambda(t)$ is ready.

\subsection{Eventual Concurrency}

Yet another subtlety of dealing with a mutable global state is a concept we'll
call eventual concurrency. Concurrency on its own broadly refers to running or
executing multiple tasks simultaneously and consistently on some shared
resources~\cite{10.5555/174770}. Eventual concurrency then refers to scenarios
where concurrency will need to be adhered to at some point in the future, but
not necessarily right now. This is probably explained most easily through
another
example.

Consider that we have element A on node 0 where $\Lambda(t)$ is up-to-date at
some time $t$, and we have element B on node 1 where $\Lambda(t)$ is
\textit{not} up-to-date at time $t$ because the mutation of $\Lambda(t)$ hasn't
been received on node 1 yet (due to Statement \ref{statement:a_before_b}).
Now suppose that element A sends a message M to element B, starting work on
element B that requires $\Lambda(t)$ to be up to date. Again, because of
Statement \ref{statement:a_before_b}, there is \textit{no guarantee} that M
arrives at element B after $\Lambda(t)$ has been mutated. It is entirely
possible to arrive before $\Lambda(t)$ is mutated on node 1, even though it was
sent after $\Lambda(t)$ was mutated on node 0.

Therefore, element B must query $\Lambda(t)$ (section
\ref{sec:async-update-lambda}) on node 1 before it does its work, even though
$\Lambda(t)$ was up-to-date on node 0 when element A sent the message. If
element B finds $\Lambda(t)$ not up-to-date, then it registers a callback. If
element B finds $\Lambda(t)$ is up-to-date, then it can continue. This necessity
of having to always query $\Lambda(t)$ when a task starts its execution,
regardless of whether $\Lambda(t)$ is up to date on the node of the sender of
the task, is what we call eventual concurrency. In practice, $\Lambda(t)$ only
needs to be queried if the task actually uses $\Lambda(t)$. If it doesn't, this
check for eventual concurrency can be skipped.

\subsection{Summary}

Previous sections went into detail about each aspect of our asynchronous
algorithm. Here, we provide a higher-level summary of the workflow for how the
time-dependent maps, their parameters, the control system, the horizon-finds, and
asynchronous communication all interact and influence each other.

\begin{itemize}
  \item The excision boundaries and the mesh (section \ref{sec:domain}) are
  controlled by time-dependent coordinate maps and their parameters $\lambda(t)$
  (section \ref{sec:time-dep-map}), which must be updated periodically
  throughout the simulation as the black holes inspiral and merge.
  \item Each $\lambda(t)$ is represented as a piecewise polynomial
  in time (Eq.~\eqref{eq:lambda}).
  \item Each piece of $\lambda(t)$ will expire (section \ref{sec:exp-time}) at
  some time and elements will have to wait for it to be mutated to continue
  their evolution past this expiration time.
  \item For these reasons, $\lambda(t)$ occupies a mutable global state in the
  simulation (section \ref{sec:global-state}).
  \item Elements must query (section \ref{sec:async-update-lambda}) $\lambda(t)$
  at a given time before they can evolve forward and eventually find a horizon.
  \item If the given time is after the expiration time, the element registers a
  callback (section \ref{sec:async-update-lambda}) for when $\lambda(t)$ is
  mutated.
  \item We mutate $\lambda(t)$ (section \ref{sec:control-error-signal}) with
  information from four previous apparent horizon-finds.
  \item Each horizon-find is done asynchronously and requires novel logic to
  handle the collection of variables from each element (section
  \ref{sec:horizon-finder}).
  \item Once four horizon-finds have happened and $\lambda^\mathrm{new}(t)$ is
  computed, $\lambda(t)$ is mutated in a threadsafe and asynchronously
  consistent manner (section \ref{sec:async-update-lambda}).
  \item Any element that registered a callback for when $\Lambda(t)$ was mutated
  is then restarted.
\end{itemize}

\subsection{Deadlocks and Errors\label{sec:deadlock}}

If any of the asynchronous algorithms described above in sections
\ref{sec:io}-\ref{sec:async-control-system} is implemented incorrectly, or if
there are edge cases that aren't taken into account, a phenomenon called a
``deadlock'' can occur. In its simplest form, a deadlock occurs when you have a
task that is waiting to receive a message but never does. Here, ``waiting'' does
not mean that a task is still allocated to a core blocking another task from
running on the core. It means that a task has finished its computation, is no
longer running on a core, and is idle until it can be restarted. A common
deadlock can happen when there are two tasks A and B, and A is waiting for a
message from B while B is waiting for a message from A. Therefore, both tasks
halt indefinitely. One scenario of a deadlock was described in section
\ref{sec:async-update-lambda} when query and mutating $\Lambda(t)$ happen
simultaneously.

The consequence of a deadlock is that every element will continue to evolve
until it no longer can. Then, once no elements can continue evolving forward in
time, the simulation will hang; i.e.~quiesce. Fortunately, Charm++ offers
quiescence detection and we have built a feature into \spectre~that uses
quiescence detection to then run additional functions that will terminate the
simulation cleanly. However, this termination will likely not happen at the time
that the user intended. Elements could be waiting to be restarted at different
times in an inconsistent state. This can be quite difficult to debug because the
simulation is no longer in the state it was when the actual error/deadlock
occurred. Additionally, it is also difficult because the simulation looks as
though it finished correctly because there was no obvious error such as an FPE
or segmentation fault. Therefore we developed a method to debug deadlocks that
outputs the deadlocked state of the simulation to disk, including copious
amounts of information on what messages each task is waiting on. This
information can be used to identify the offending task that caused the
deadlock.
  
We'll use the horizon interpolator as an example of the information that is
output after a deadlock since it couples many different communication patterns.
First, we output which elements have sent their volume variables to the horizon
interpolator at given times. If a horizon-find did not complete at a certain
time, and the horizon interpolator is missing an element's variables, then we
know we need to look at why the horizon interpolator never received the
variables. This could be because the element never sent a message, the message
got lost, or the message did arrive but the logic in the horizon interpolator
didn't handle the message properly. We also output which horizon-finds have
completed, are currently being worked on, or are waiting to be worked on, which
can help us determine how far forward in time some elements have evolved.
Additionally, we output the points of the current fast-flow iteration trial
surface and which points already have been interpolated to. This can tell us if
interpolation failed for a certain trial point or if some elements haven't sent
their variables. Furthermore, we also output the nodes/cores that elements sent
their variables from, which can be helpful in determining if the problem is with
inter-node messages.

Another useful diagnostic for debugging a deadlock is to output $\Lambda(t)$,
whether any callbacks were registered, and which elements
those callbacks would have restarted. This can usually be used in
conjunction with other deadlock information (such as the horizon interpolator)
to determine if an element simply couldn't evolve forward because of an
expiration time or because it was waiting for other data necessary for its
evolution. It can also be used to ensure that an element \textit{was} restarted
and thus direct us to other areas to look for the bug.

\section{Application to binary neutron stars\label{sec:bns}}

Even though the techniques described in previous sections were developed for use
in black hole simulations, many of them can still be applied to simulations of
other systems that are done with asynchronous parallelism. Here we describe how
they can be applied to simulations of binary neutron stars (BNSs) as they are
done in \spectre~\cite{Deppe:2024ckt}.

The surface of a neutron star is a discontinuity in the matter profile and is
usually handled with shock capturing methods \cite{RezzollaBook}. If the
computational domain is stationary and a binary neutron star (BNS) system is
orbiting, a NS surface will pass by a grid point on a timescale of roughly
\begin{equation}
  \tau_{\mathrm{stationary}} \propto \frac{\Delta x}{\Omega d/2},
\end{equation}
where $\Delta x$ is the spacing between grid points, $d$ is the separation
between the NSs, and $\Omega$ is the orbital angular velocity. However, if the
computational grid is rotating at the same rate $\Omega$ as the stars are
orbiting, the surface of the star would move between elements much more slowly.
Having the computational domain move with the fluid can limit numerical
diffusion (e.g. \cite{Mandal_2023}). If we boost into this corotating frame, the
only movement of the stars would be radially inward as they inspiral. They will
pass by grid points on a timescale of roughly
\begin{equation}
  \tau_{\mathrm{corotating}} \propto \frac{\Delta x}{\dot{r}},
\end{equation}
where $\dot{r}$ is the radial velocity of the NSs. If we use $\Delta x =
200\,\mathrm{m}$, $d = 32.4\,\mathrm{km}$, $\Omega =
1.1\times10^3\,\mathrm{rad/s}$, and $\dot{r} = 11\, \mathrm{km/s}$ (taken from
\cite{Habib:2025inprep}), we have
\begin{equation}
  \tau_{\mathrm{stationary}} \approx 0.01\, \mathrm{ms} \ll
  \tau_{\mathrm{corotating}} \approx 20\, \mathrm{ms},
\end{equation}
showing how much more slowly the discontinuity of the NS would move through a
corotating grid.

{
\begin{figure}
\includegraphics[width=0.48\linewidth]{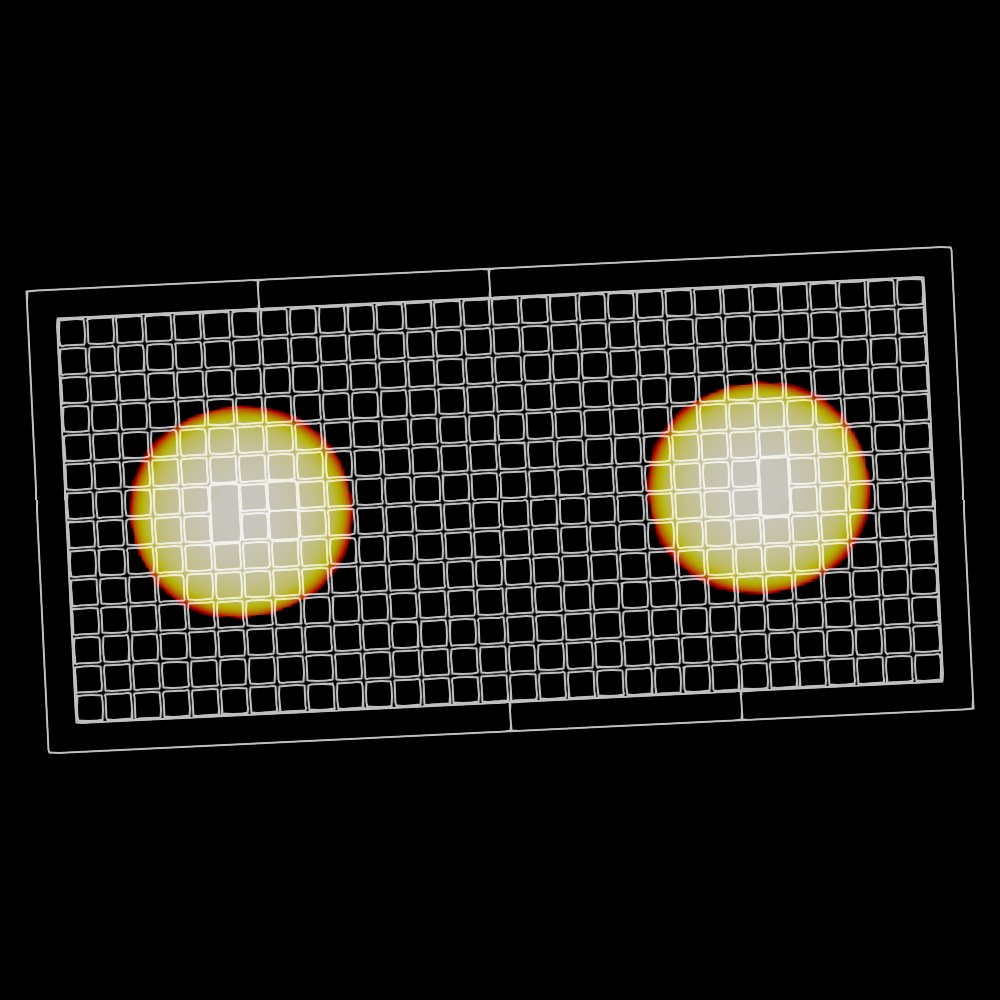}
\includegraphics[width=0.48\linewidth]{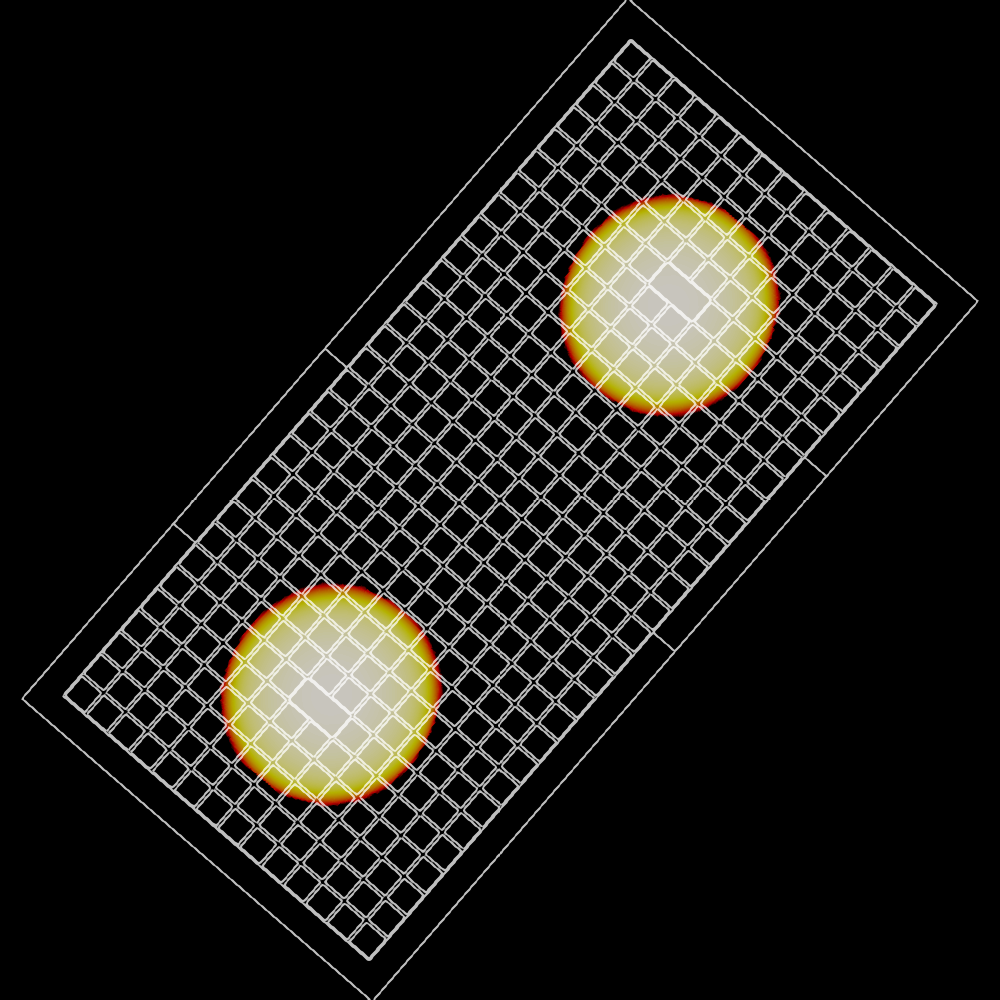} \\[1ex]
\includegraphics[width=0.48\linewidth]{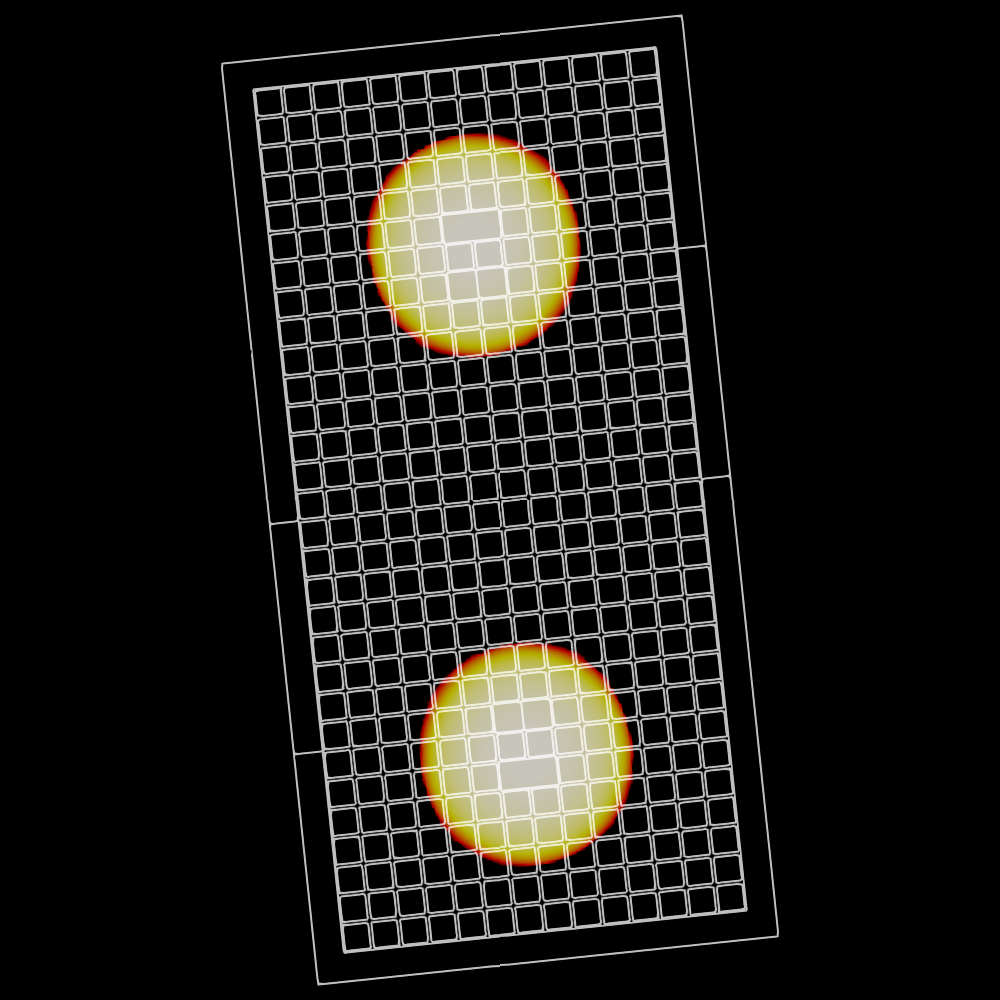}
\includegraphics[width=0.48\linewidth]{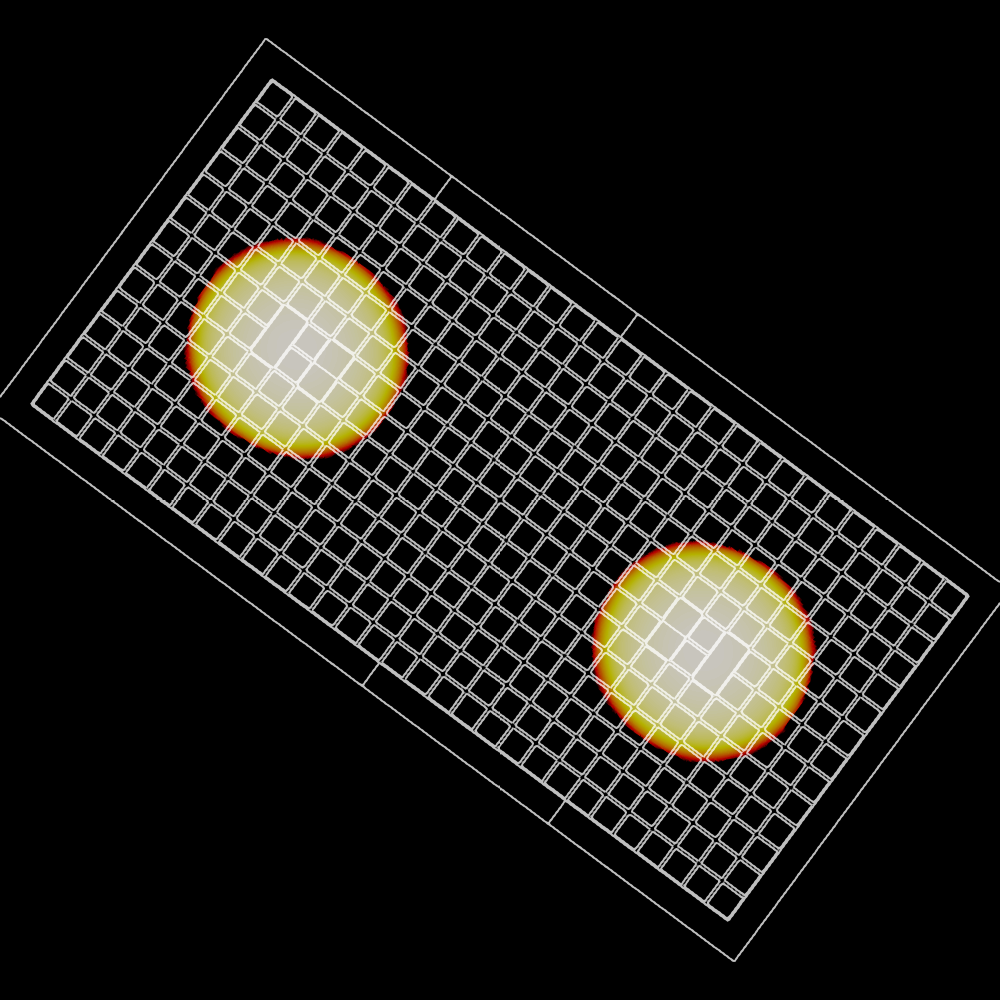}
\caption{Snapshots of the rest mass density and grid of a BNS system where the
computational domain is rotated and controlled using the methods in this
paper.}
\label{fig:bns} 
\end{figure}
}

To rotate the computational domain with the orbit of the NSs, though, we don't
\textit{a priori} know the evolution of the orbital velocity of the BNSs, just
like we also cannot know it for BBHs. Therefore, we need a dynamic way of
tracking the evolution of the BNS system. The overall method in section
\ref{sec:async-control-system} can be applied here to dynamically track and
update the angular velocity of the computational domain, except we don't use
horizon-finds as our measurements. Instead, we find the center of mass of each
NS and the system as a whole and use these measurements to compute $Q(t)$, its
derivatives, $U(t)$, and eventually $\lambda(t)$. In figure \ref{fig:bns}, we
show snapshots of an asynchronous BNS simulation done with \spectre~where the
NSs are tracked with the methods in this paper \cite{Habib:2025inprep}.

\section{Conclusion\label{sec:conclusion}}

In this paper, we present novel methods for finding and tracking apparent
horizons in asynchronous simulations of black holes using a feedback control
system. We first introduce asynchronous parallelism in section \ref{sec:async}
and provide a Statement \ref{statement:a_before_b} about asynchronous
parallelism that is the cornerstone of all our algorithms. Additionally, we
present an example of how a common algorithm like I/O must be rethought when
implementing it in an asynchronous application compared to a synchronous one. We
then detail how our asynchronous horizon finding algorithm works compared to a
synchronous implementation in section \ref{sec:horizon-finder}. Maybe
surprisingly, implementing an asynchronous horizon finder itself is not the most
complicated parallel aspect for our dynamic tracking of apparent horizons.

We then present details for a synchronous feedback control system in section
\ref{sec:sync-control-system} and its asynchronous counterpart in section
\ref{sec:async-control-system}. This feedback control system dynamically adjusts
the time-dependent coordinate mappings used to move and distort the
computational domain so the position and shape of the excision boundaries match
those of the apparent horizons. The critical conclusion of this work is that the
time-dependent coordinate mappings act as shared mutable global states in the
asynchronous simulation. We describe the extreme care that needs to be taken
when dealing with these shared mutable global states to avoid deadlocks, errors,
and inconsistencies in an evolution. All these algorithms have been added to the
open-source numerical relativity code \spectre~and have already been used in a
number of publications
~\cite{Deppe:2024ckt, Lovelace:2024wra, Lara:2024rwa, Wittek:2024pis,
  Lara:2025kzj, Habib:2025inprep}.

Additionally, in section \ref{sec:bns} we show that, omitting horizon finding,
the asynchronous feedback control system can be used for evolutions of BNSs to
rotate the computational domain to follow the orbit of the NSs. This shows the
versatility of these algorithms and their potential to be applied to other
evolutions where dynamic tracking of a feature would be useful. As future work,
we plan to apply the methods in this paper to other problems, such as black
hole-neutron star binaries or utilizing \spectre's flexible domains to track
features in other simulations of general relativistic magnetohydrodynamic
systems. Additionally, we plan to perform detailed analyses to ascertain the
explicit performance improvement of these asynchronous algorithms over
synchronous ones, particularly the scalability compared a code such as \spec.

Though we did not do a performance analysis in this work, the algorithms
discussed here were developed in large part because it is believed that
asynchronous algorithms will significantly speed up numerical relativity
simulations as the number of processors available on machines continues to rise
(e.g., the current state-of-the art BBH code \texttt{SpEC} can spend a
significant fraction of its time in \texttt{MPI\_Wait} calls for high-mass ratio
and high-spin simulations). Results in~\cite{Lovelace:2024wra} show strong
promise that asynchronous methods will help to outperform synchronous ones for
BBH simulations. We believe that asynchronous parallelism and the methods
described here will be a critical ingredient in high-performance BBH
simulations.

\acknowledgements

KCN would like to thank Geoffrey Lovelace for running \spectre~BBH simulations
and finding various deadlocks that needed to be fixed and accounted for in the
algorithms presented in this work. This material is based upon work supported by
the National Science Foundation under Grants No.~PHY-2309211, No.~PHY-2309231,
and No.~OAC-2209656 at Caltech; %
by No.~PHY-2407742, No.~PHY-2207342, and No.~OAC-2209655 at Cornell; and by
No.~PHY-2208014, No.~PHY-1606522 and AST-2219109 at Cal State Fullerton. Any
opinions, findings, and conclusions or recommendations expressed in this
material are those of the author(s) and do not necessarily reflect the views of
the National Science Foundation. This work was supported by the Sherman
Fairchild Foundation at Caltech and Cornell, and by Nicholas and Lee Begovich
and the Dan Black Family Trust at Cal State Fullerton.
Charm++/Converse~\cite{laxmikant_kale_2020_3972617} was developed by the
Parallel Programming Laboratory in the Department of Computer Science at the
University of Illinois at Urbana-Champaign. The figures in this article were
produced with \texttt{TikZ}~\cite{tikz} and
\texttt{ParaView}~\cite{paraview, paraview2}. Computations were performed with
the Wheeler cluster and the Resnick High Performance Computing Center at
Caltech, the mbot cluster at Cornell, and the Ocean cluster at Cal State
Fullerton.

\section*{References}
\bibliographystyle{unsrt}
\bibliography{refs}
\end{document}